\documentstyle{camera}
\newcommand{\AmS}{{\protect\the\textfont2
  A\kern-.1667em\lower.5ex\hbox{M}\kern-.125emS}}

\newcommand{\beq}{\begin{equation}}
\newcommand{\eeq}{\end{equation}}
\newcommand{\bea}{\begin{eqnarray}}
\newcommand{\eea}{\end{eqnarray}}

\def\dm2{\Delta m^2}
\def\sq2{sin^2(2\Theta)}

\input epsf

\begin{document}

%%%%%%%%%%%%%%%%%%%%%%%%%%%%%%%%%%%%%%%%%%%%%%%%%%%%%%%%
% The title, all uppercase; if you want to split it in
% two or more lines, put a \\ macro at each line break
% example:
%   \title{TITLE: FIRST LINE\\ SECOND LINE}
%
\title{PHYSICS RESULTS FROM THE ARGO-YBJ EXPERIMENT}

%%%%%%%%%%%%%%%%%%%%%%%%%%%%%%%%%%%%%%%%%%%%%%%%%%%%%%%%
% The Author(S), Separated By Commas; Do not put a
% comma before the last author, use instead the \And
% macro which produces a normal ``and'' in the
% caps/small caps context
%
\author{GIUSEPPE DI SCIASCIO \\ on behalf of the ARGO-YBJ collaboration}

%%%%%%%%%%%%%%%%%%%%%%%%%%%%%%%%%%%%%%%%%%%%%%%%%%%%%%%%
%
\organization{INFN, Sez. Roma TorVergata\\ Via della Ricerca
Scientifica 1, I-00133 Roma, Italy}

\maketitle

\begin{abstract}
The ARGO-YBJ experiment has been in stable data taking since
November 2007 at the YangBaJing Cosmic Ray Laboratory (Tibet, P.R.
China, 4300 m a.s.l.). In this paper we report a few selected
results in Gamma-Ray Astronomy (Crab Nebula and Mrk421
observations, search for high energy tails of GRBs) and Cosmic Ray
Physics (Moon and Sun shadow observations, proton-air cross
section and $\overline{p}$/p preliminary measurements).
\end{abstract}
\vspace{1.0cm}

\section{The detector}

The ARGO-YBJ experiment, located at the YangBaJing Cosmic Ray
Laboratory (Tibet, P.R. China, 4300 m a.s.l.), is an air shower
array exploiting the full coverage approach at very high altitude,
with the aim of studying the cosmic radiation with a low energy
threshold (a few hundreds of GeV). The detector is constituted by
a central carpet, made of a single layer of Resistive Plate
Chambers (RPCs) with $\sim$92$\%$ of active area, enclosed by a
guard ring partially instrumented ($\sim$40$\%$) mainly to improve
the rejection capability for external events. The apparatus has a
modular structure, the basic element being a cluster
(5.7$\times$7.6 m$^2$), divided into 12 RPCs (2.8$\times$1.25
m$^2$ each). Each chamber is read by 80 strips of 6.75$\times$61.8
cm$^2$ (the spatial pixel), logically organized in 10 independent
pads of 55.6$\times$61.8 cm$^2$ which are individually acquired
and represent the time pixel of the detector. In order to extend
the dynamic range up to PeV energies, a charge read-out has been
implemented by instrumenting every RPC also with two large size
pads of dimension 140$\times$125 cm$^2$ each. The full detector is
composed of 154 clusters for a total active surface of $\sim$6700
m$^2$ (Aielli G. et al., 2006). A 0.5 cm thick lead converter will
cover the apparatus uniformly in order to improve the angular
resolution at lower energies.

%The main features of the ARGO-YBJ experiment are: (1) time
%resolution $\sim$1 ns; (2) space information from strips; (3) time
%information from pads. Due to its small size pixels, the detector
%is able to image the shower profile with an unprecedented
%granularity, with high duty cycle ($> 90\%$) in the typical field
%of view of an EAS array ($\sim$2 sr).
%

The detector is connected to two different data acquisition
systems, working independently, and corresponding to the two
operation modes, shower and scaler. In shower mode, for each event
the location and timing of every detected particle is recorded,
allowing the lateral distribution and arrival direction
reconstruction (Di Sciascio G. et al., 2007). In scaler mode the
total counts are measured every 0.5 s, with very poor information
on both the space distribution and arrival direction of the
detected particles. For each cluster, the signal coming from the
120 pads is added up and put in coincidence in a narrow time
window (150 ns), giving the counting rates of $\ge$1, $\ge$2,
$\ge$3, $\ge$4 pads, that are read by four independent scaler
channels. The corresponding measured rates are, respectively,
$\sim$40 kHz, $\sim$2 kHz, $\sim$300 Hz and $\sim$120 Hz for each
cluster (Aielli G. et al., 2008).

Since 2007 November the detector has been in stable data taking
with a multiplicity trigger N$_{pad}\geq$20 and a duty cycle $\geq
85\%$: the trigger rate is about 3.5 kHz. In this paper we report
on the first results in Gamma-Ray Astronomy and Cosmic Ray
Physics.

\section{Gamma-Ray Astronomy}

In this analysis all events
%
%with $\chi^2<$50 (resulting from the temporal profile fitting)
%
are considered, without any internal shower selection. No lead
converter has been mounted on the RPCs yet (with the lead we
expect an improvement of the angular resolution and, consequently,
of the sensitivity, of $\approx$40$\%$ at TeV energies). The
cosmic ray background rejection is performed by exploiting the
good angular resolution of the detector and the fact that at this
altitude the trigger efficiency of TeV $\gamma$-rays is $\approx$2
times larger than that of protons at fixed energy. No additional
$\gamma$/hadron discrimination algorithms have been yet applied.
We expect an improvement of our sensitivity of a factor
$\approx$1.5 by applying "topological" selection criteria
presently under study (Dattoli M. et al., 2007).

\subsection{Angular Resolution}

%%%%%%%%%%%%%%%%%%%%%%%%%%%%%%%%%%%%%%%%%%%%%%%%%%%%%%%%%%%%%%%%%%%%
\begin{figure}[t!]
\begin{minipage}[t]{.5\linewidth}
\begin{center}
\vspace{0.9cm} \epsfysize=4.5cm \hspace{0.3cm}
\epsfbox{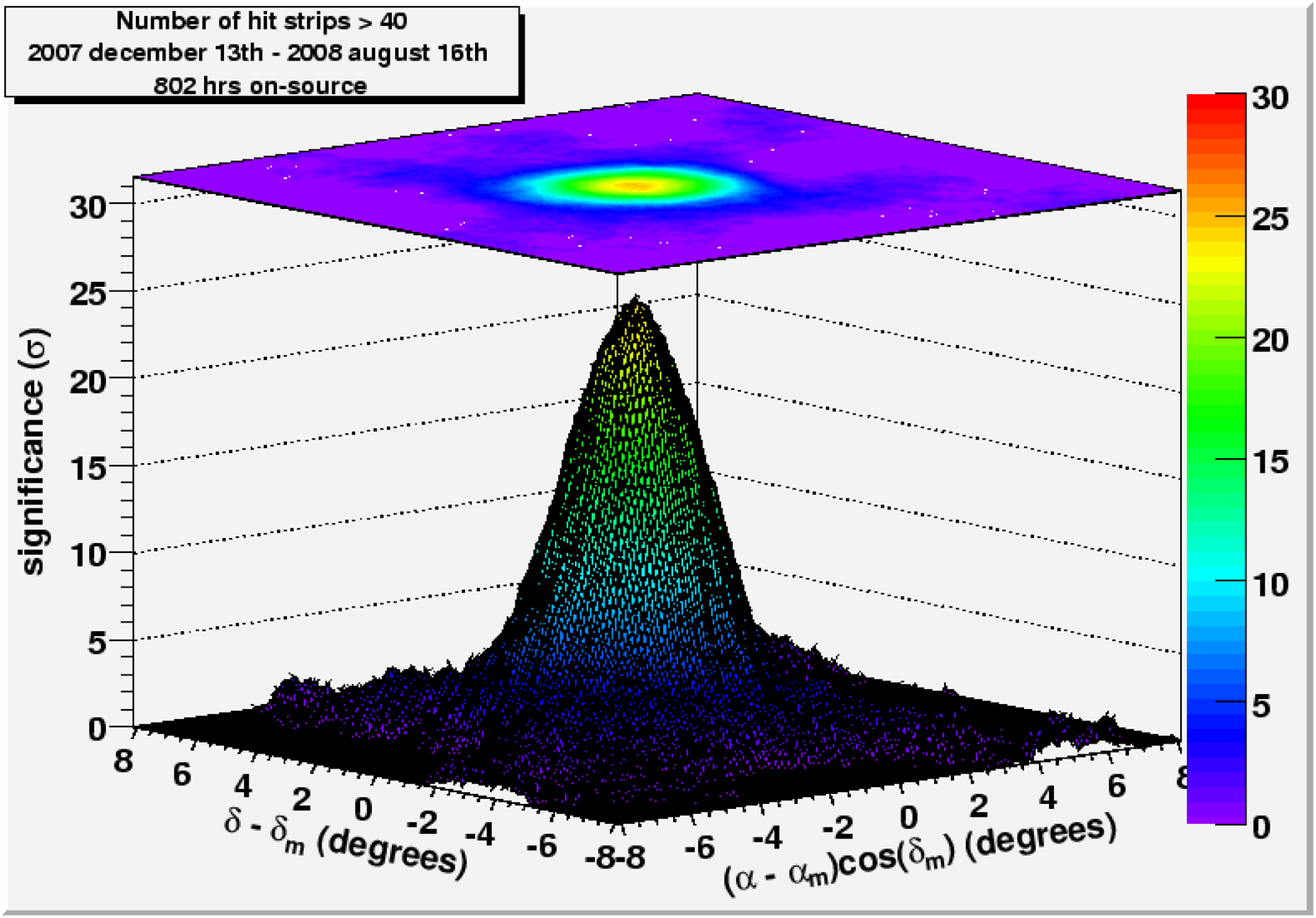}
  \end{center}
\end{minipage}\hfill
\begin{minipage}[t]{.47\linewidth}
  \begin{center}
\epsfysize=6.cm \hspace{0.5cm}
\epsfbox{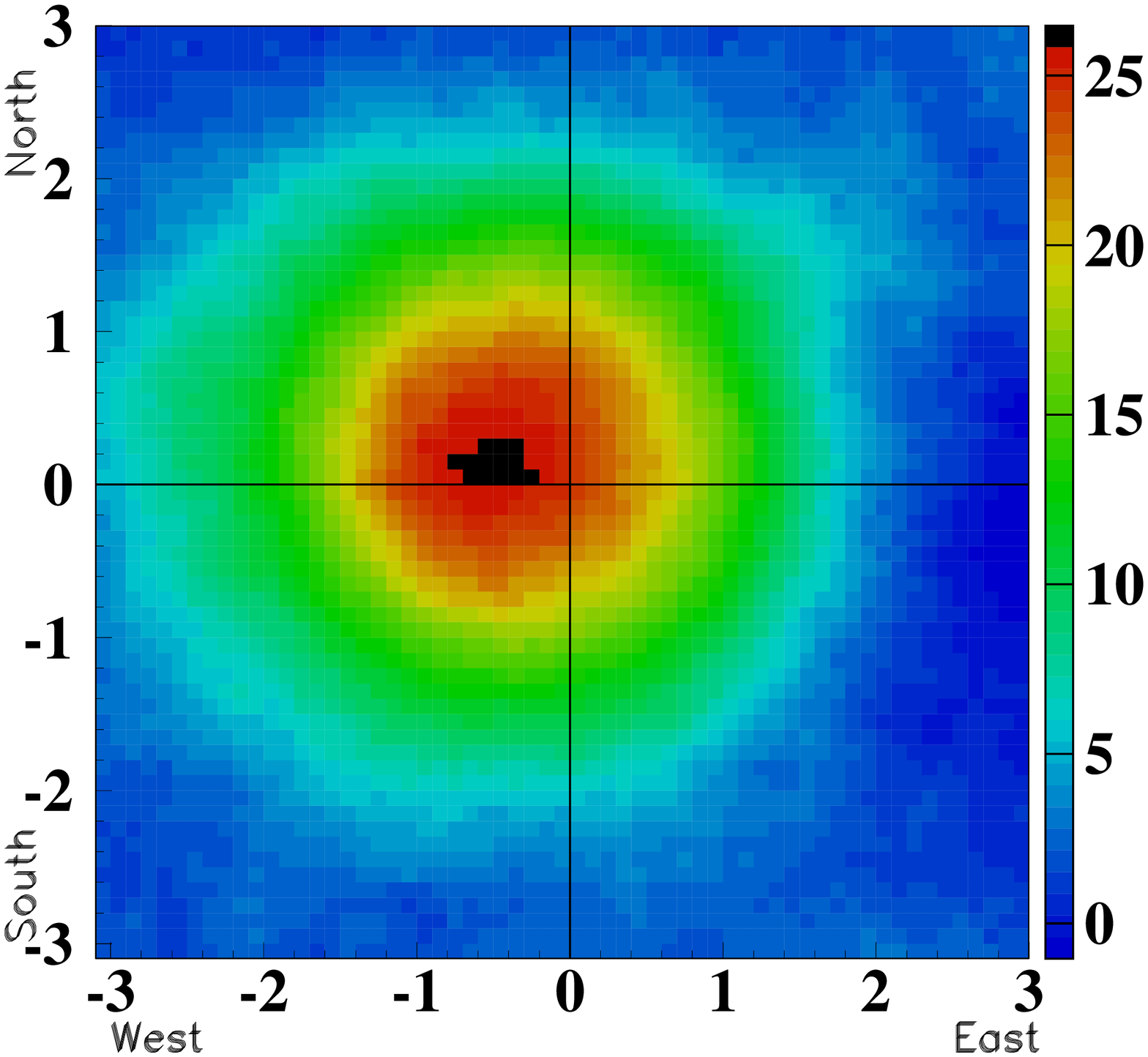} \vspace{0.3cm}
  \end{center}
\end{minipage}\hfill
\vspace{-0.7cm} \caption[h]{Point Spread Function of the ARGO-YBJ
detector around the Moon observed in 2008 for showers with
N$_{pad}\geq$40 (left plot). In the right plot the corresponding
Moon shadow significance map is shown. The color scale gives the
significance.} \label{fig:moon1}
\end{figure}
%%%%%%%%%%%%%%%%%%%%%%%%%%%%%%%%%%%%%%%%%%%%%%%%%%%%%%%%%%%%%%%%%%%%
%

%
%%%%%%%%%%%%%%%%%%%%%%%%%%%%%%%%%%%%%%%%%%%%%%%%%%%%%%%%%%%%%%%%%%%%
\begin{figure}[t!]
\begin{minipage}[t]{.47\linewidth}
\begin{center}
\epsfysize=6.5cm \hspace{0.5cm}
\epsfbox{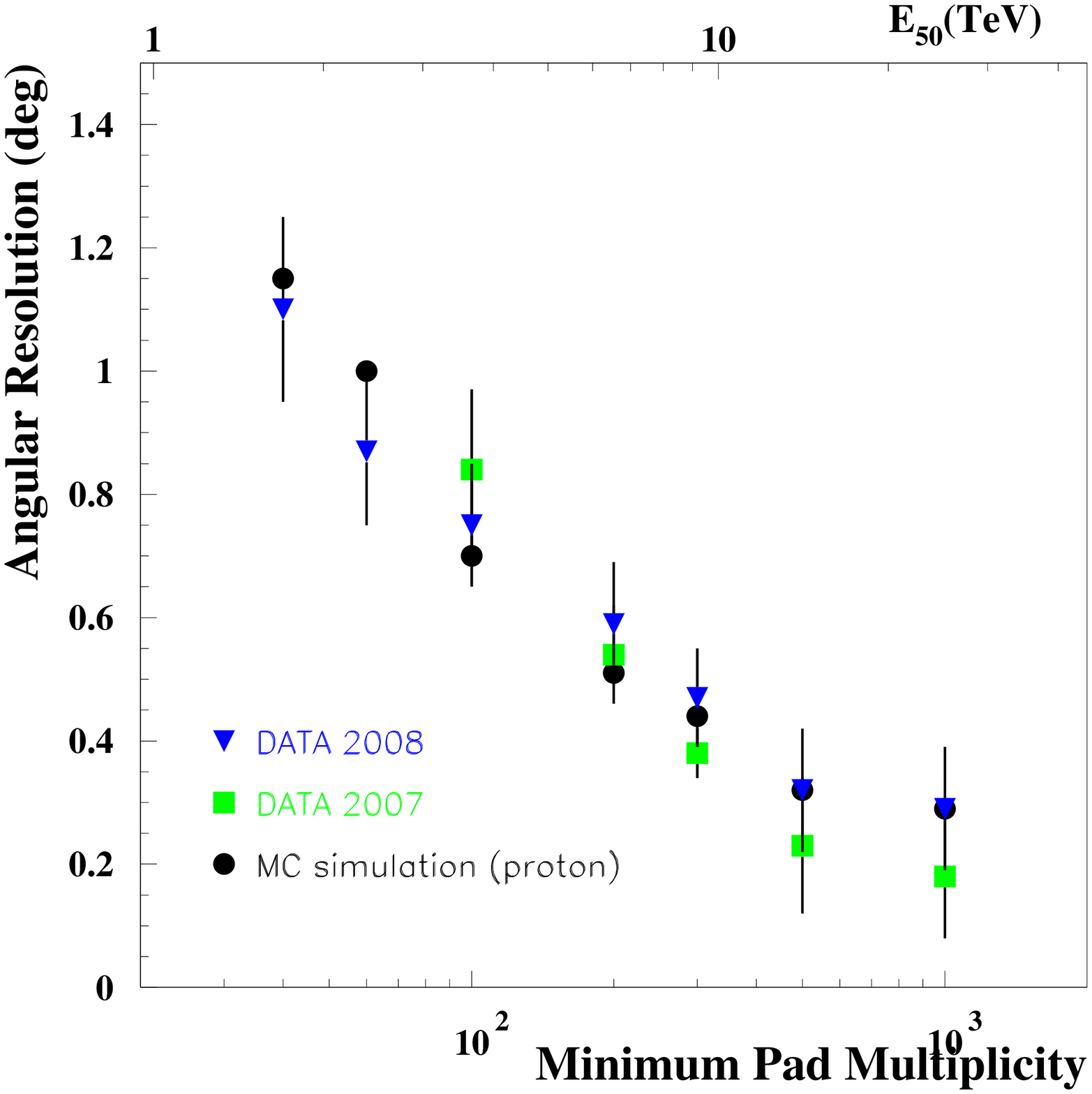} \vspace{-0.5cm}
\caption[h]{Measured angular resolution compared with MC
simulations as a function of pad multiplicity.}
\label{fig:angresol}
  \end{center}
\end{minipage}\hfill
\begin{minipage}[t]{.47\linewidth}
  \begin{center}
\epsfysize=6.5cm \hspace{0.5cm}
\epsfbox{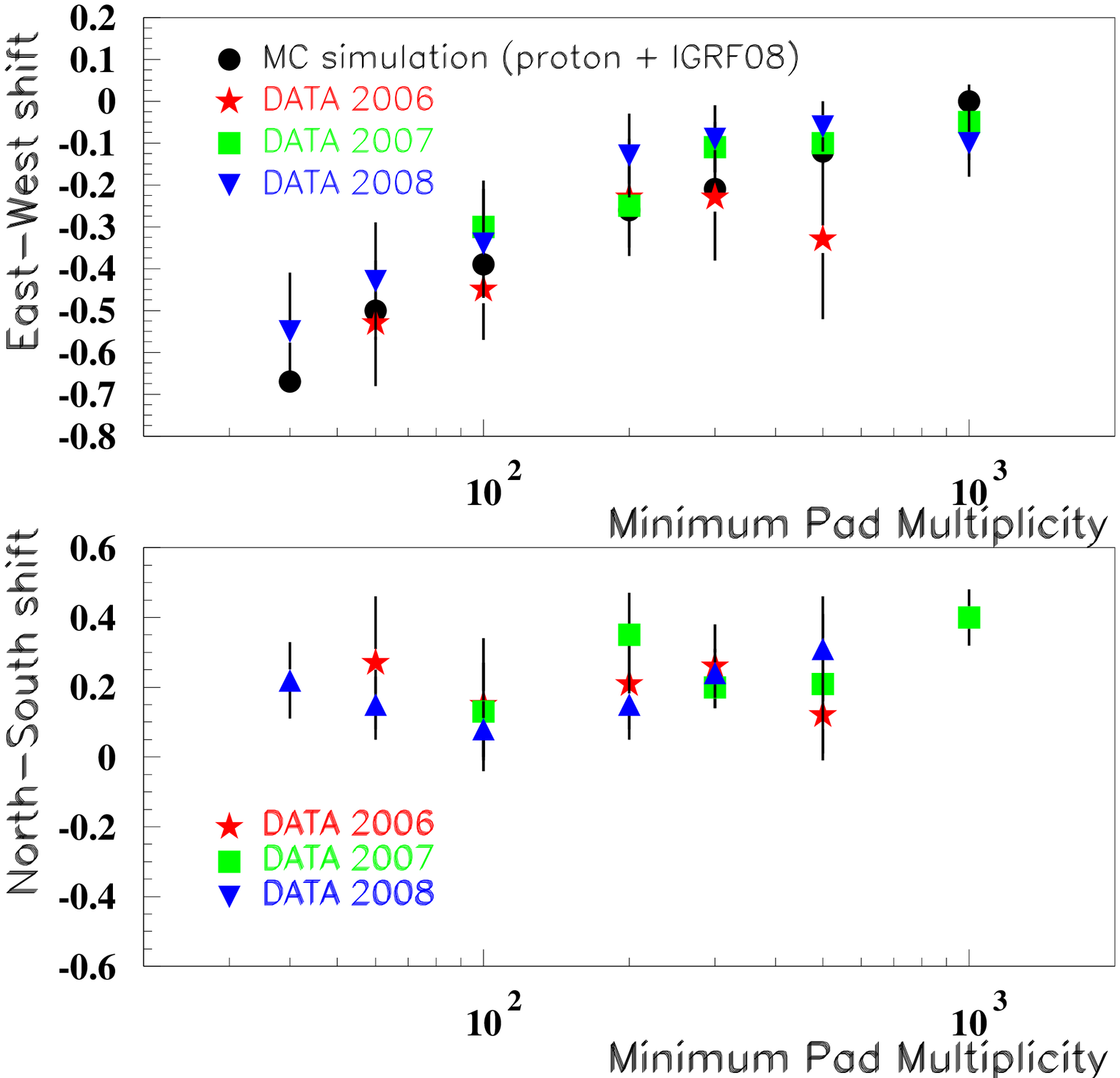} \vspace{-0.5cm}
\caption[h]{Displacement of the Moon shadow in the East-West
(upper panel) and North-South (lower panel) directions as a
function of pad multiplicity.} \label{fig:moonshift}
  \end{center}
\end{minipage}\hfill
\end{figure}
%%%%%%%%%%%%%%%%%%%%%%%%%%%%%%%%%%%%%%%%%%%%%%%%%%%%%%%%%%%%%%%%%%%%
%

We have measured the pointing accuracy of the ARGO-YBJ detector by
studying the Moon shadow. Cosmic rays are hampered by the Moon,
therefore a deficit of cosmic rays in its direction is expected
(the so-called "Moon shadow"). The Moon shadow is an important
tool to calibrate the performance of an air shower array. In fact,
the size of the deficit allows a measurement of the angular
resolution and the position of the deficit allows the evaluation
of the absolute pointing accuracy of the detector. In addition,
positively charged particles are deflected towards East due to the
geomagnetic field by an angle $\Delta\theta\sim
1.6^{\circ}/E(TeV)$. Therefore, the observation of the
displacement of the Moon provides a direct check of the relation
between shower size and primary energy thus calibrating the
detector.

The ARGO-YBJ experiment is observing the Moon shadow with a
sensitivity of about 10 standard deviations per month at a
multiplicity N$_{pad}\geq$40, with the zenith angle
$\theta<50^{\circ}$ (corresponding to a proton median energy
E$_{50}\approx$1.8 TeV). In Fig.\ref{fig:moon1} the shadow
observed in the period December 2007 - August 2008 (802 hours
on-source) is shown. In the left plot the 3D view of the shadow
(i.e., the Point Spread Function of the detector) is shown. The
sensitivity of the observation is about 26 standard deviations as
can be seen from the Moon shadow map in the right plot of
Fig.\ref{fig:moon1}. In Fig.\ref{fig:angresol} the angular
resolution measured with the Moon shadow method in 2007 and 2008
is compared to MC simulation. The upper scale refers to the median
energy of the triggered protons. As can be seen, the values are in
fair agreement: the angular resolution of the ARGO-YBJ experiment
is less than 0.6$^{\circ}$ for N$_{pad}\geq$200 (E$_{50}\approx$6
TeV). In Fig.\ref{fig:moonshift} the displacements of the Moon
shadow in the East-West (upper panel) and North-South (lower
panel) directions are shown. In the upper figure the measured
shift towards West due to the geomagnetic field is compared to a
MC simulation of the proton propagation in Earth-Moon system. As
can be seen, the fair agreement between the measurements and the
simulations make us confident about the energy calibration of the
detector. The study of the displacement along the North-South
axis, not affected by the geomagnetic field at the Yangbajing
site, enables us to estimate the magnitude of the systematic
pointing error without the Moon shadow simulation. From this
preliminary analysis it results that there is a residual
systematic shift towards North of $\approx$0.2$^{\circ}$, less
than the angular resolution. Deeper analysis is in progress in
order to eliminate this pointing error. I would like to emphasize
that these results are stable even in periods mainly devoted to
installation and debugging operations (2006 and 2007). This makes
us confident about the stability of the detector.

The shadow of the Sun measured in the period December 2007 -
August 2008 (954 hours on-source with $\theta<50^{\circ}$) for
events with N$_{pad}\geq$40 is shown in Fig. \ref{fig:sun08}. The
significance of the maximum event deficit is about 25$\sigma$: the
shadow is well visible, as expected near the solar minimum. The
displacement of the shadow from the apparent position of the Sun
could be explained by the effect of the geomagnetic field and of
the solar and Interplanetary Magnetic Fields (IMF) whose
configuration considerably changes with the phases of the solar
activity cycle (Amenomori M. et al., 2000). Therefore, high energy
cosmic rays may bring direct information on the structure of the
IMF under the influence of the solar activity. This result shows
the capability of the ARGO-YBJ experiment to study the effect of
the IMF on the Sun shadow down to the TeV energy region with a
monthly recurrence.

\subsection{Gamma Ray Sources}

%%%%%%%%%%%%%%%%%%%%%%%%%%%%%%%%%%%%%%%%%%%%%%%%%%%%%%%%%%%%%%%%%%%%
\begin{figure}[t!]
\begin{minipage}[t]{.47\linewidth}
\begin{center}
\epsfysize=6.5cm \hspace{0.5cm}
\epsfbox{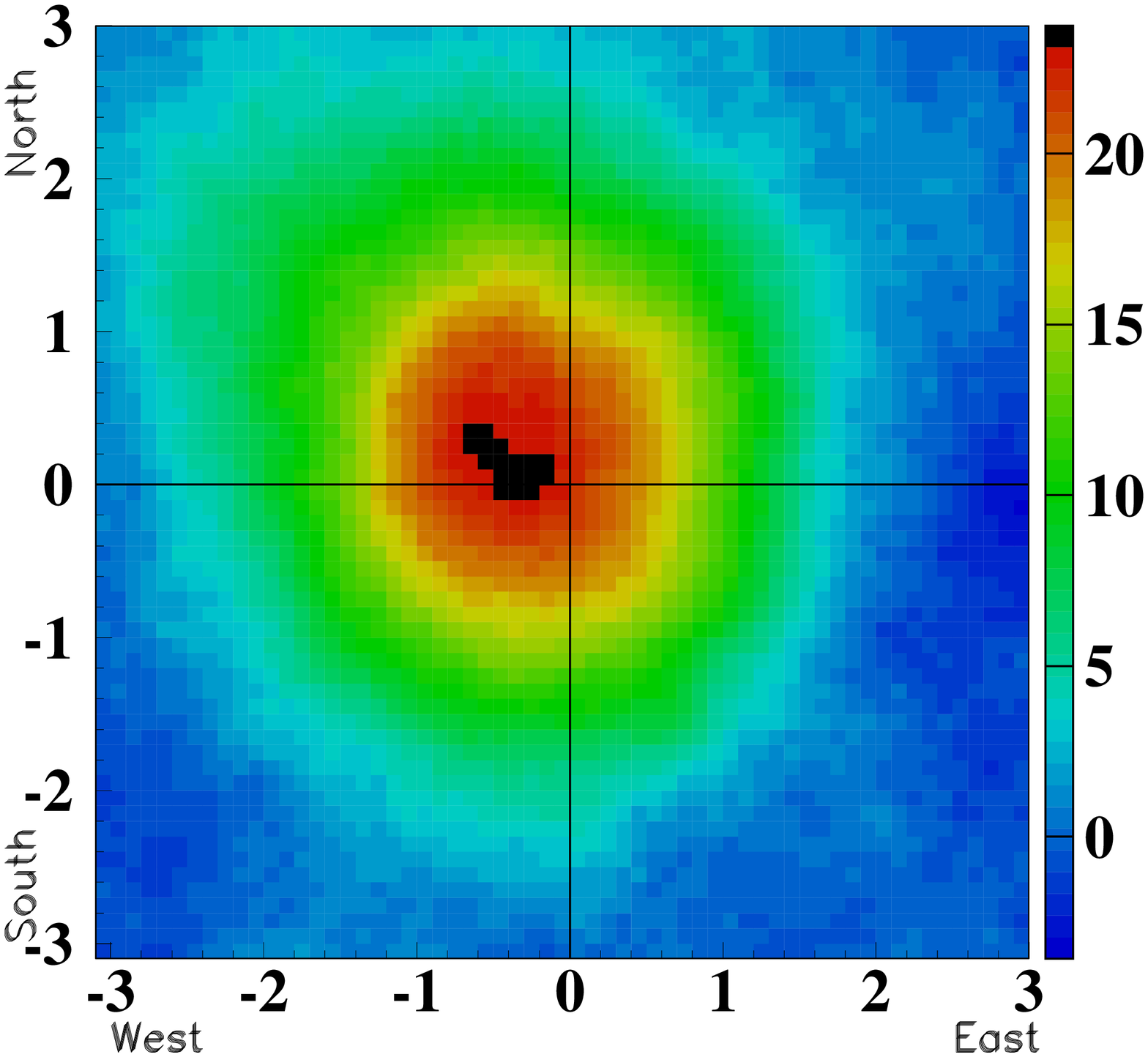} \vspace{-0.5cm}
\caption[h]{Shadow of the Sun observed from December 2007 to
August 2008.} \label{fig:sun08}
  \end{center}
\end{minipage}\hfill
\begin{minipage}[t]{.47\linewidth}
  \begin{center}
\epsfysize=6.5cm \hspace{0.5cm}
\epsfbox{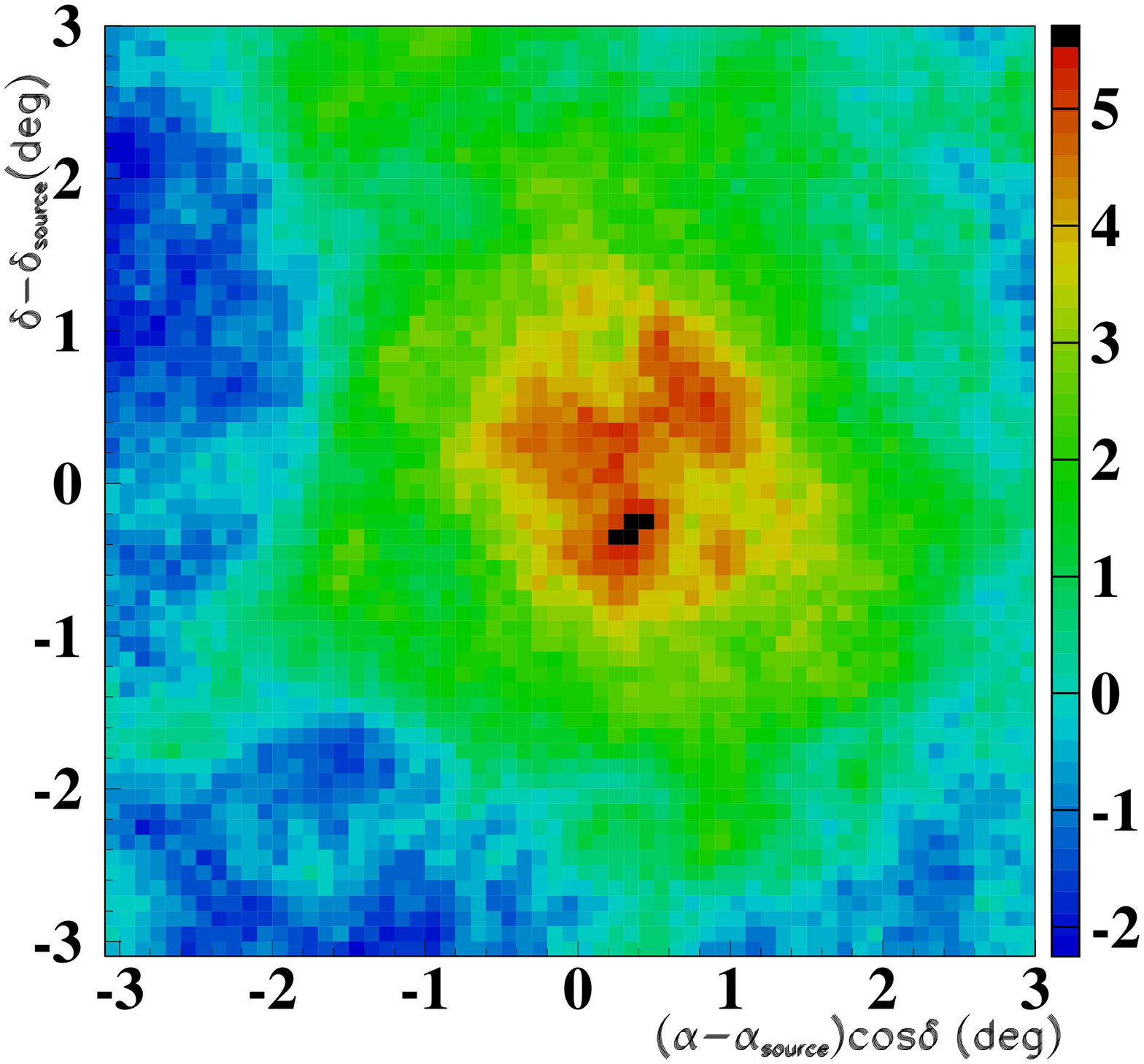} \vspace{-0.5cm}
\caption[h]{Significance map of the Crab Nebula region observed
from January to August 2008.} \label{fig:crab08}
  \end{center}
\end{minipage}\hfill
\end{figure}
%%%%%%%%%%%%%%%%%%%%%%%%%%%%%%%%%%%%%%%%%%%%%%%%%%%%%%%%%%%%%%%%%%%%
%

Fig.\ref{fig:crab08} shows the significance map of the Crab Nebula
region measured with the full ARGO-YBJ detector during days 1-229
of 2008 (1143 hours on-source with $\theta<40^{\circ}$) with a
multiplicity N$_{pad}\geq$40, corresponding to a photon median
energy of about 1 TeV (E$_{mode}\sim$0.4 TeV). As can be seen from
the map a clear signal is evident at a level of about 6 standard
deviations.
%
%We also observed a signal from the Crab Nebula at about 5 standard
%deviations level in the days 24 - 110 of 2007 (241 hours
%on-source) with a multiplicity N$_{pad}\geq$100, corresponding to
%a photon median energy of about 2.4 TeV. In this period only the
%central carpet was in data taking.
%
Deeper analysis and further data integration are in progress in
order to evaluate the final sensitivity of the detector.
%
% and to remove the residual pointing error.
%
Nevertheless, we note that this is the first time that an air
shower array is able to detect photons from a point source with
such a low peak energy.

The detector started recording data with the full central carpet
during the X-ray flare of Mrk421 in July 2006. The ARGO-YBJ
experiment observed this source in the days 190 - 245 (109 hours
on-source) with a multiplicity N$_{pad}\geq$60. A clear evidence
of a TeV emission at a level of about 6 standard deviations was
observed. The detector was in its commissioning phase, therefore
new analysis are in progress to properly evaluate the statistical
significance of the observation.

\begin{figure}[t!]
\begin{minipage}[t]{.47\linewidth}
\begin{center}
\epsfysize=6.5cm \hspace{0.5cm}
\epsfbox{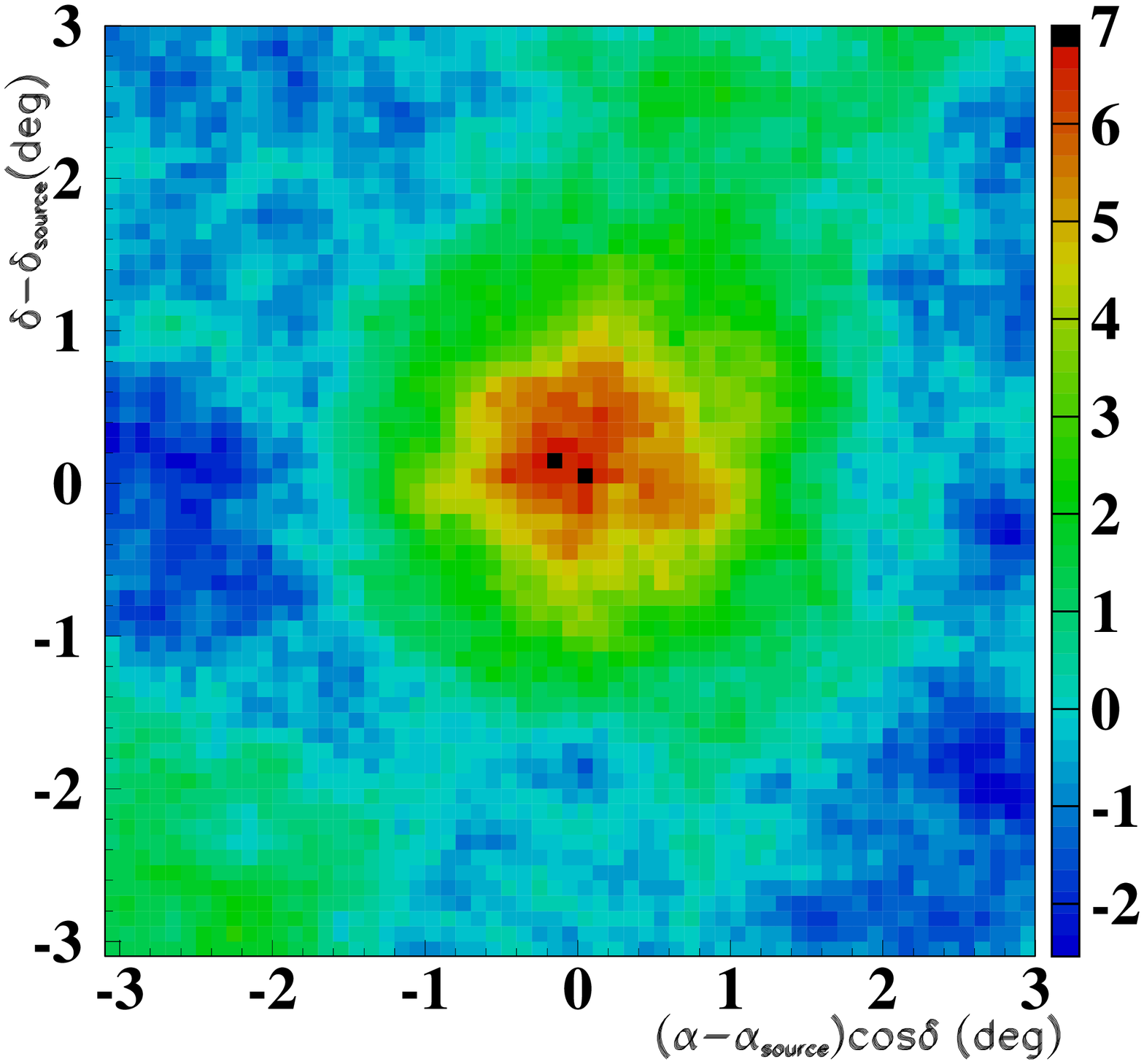} \vspace{-0.5cm}
\caption[h]{Significance map of the Mrk421 region observed up to
August 2008.} \label{fig:mrk421_08}
  \end{center}
\end{minipage}\hfill
\begin{minipage}[t]{.47\linewidth}
  \begin{center}
\epsfysize=6.5cm \hspace{0.5cm}
\epsfbox{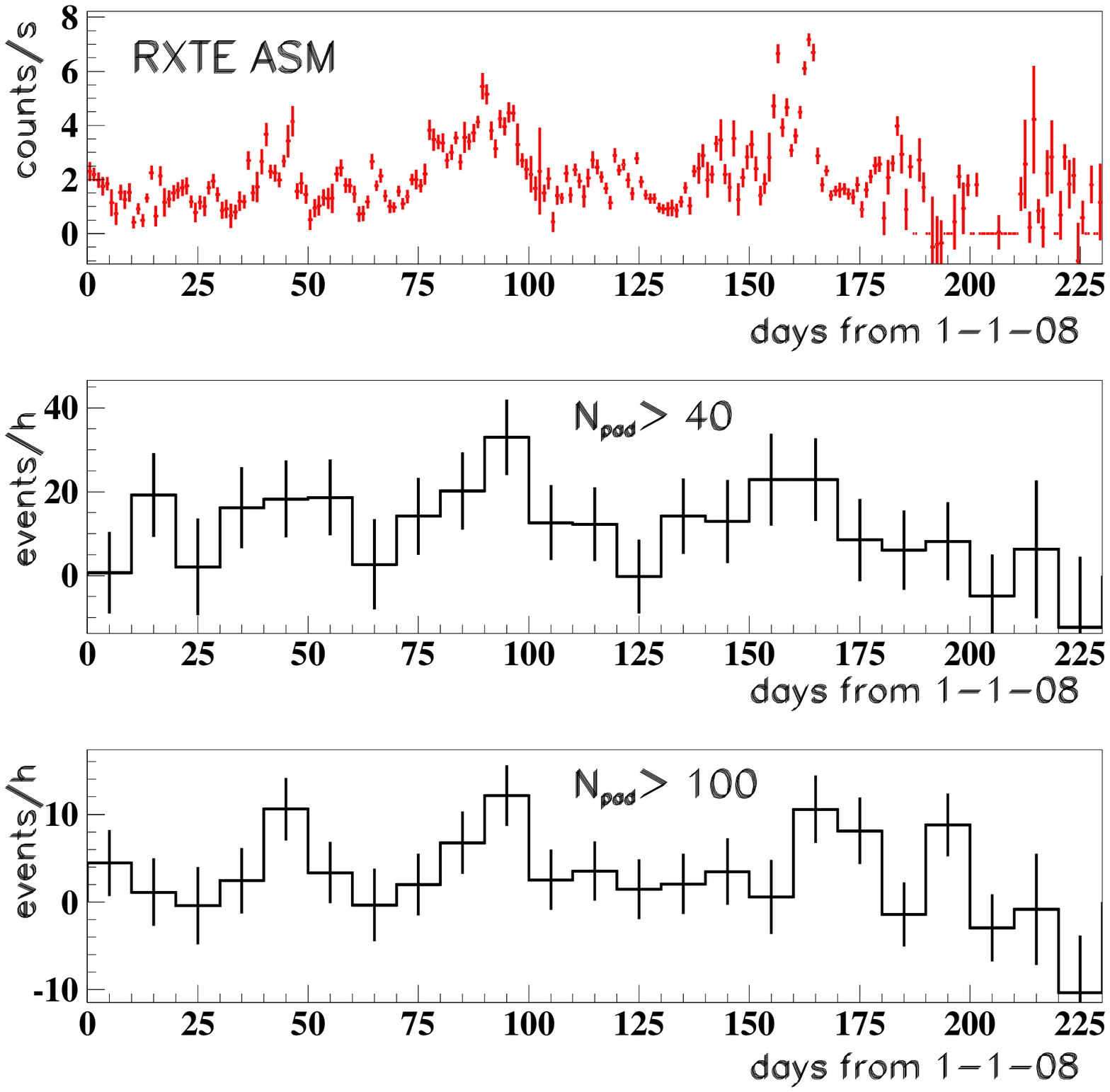} \vspace{-0.5cm}
\caption[h]{Correlation between X-ray data and TeV photons for two
multiplicity values.} \label{fig_mrk421_xray}
  \end{center}
\end{minipage}\hfill
\end{figure}

Mrk421 was again flaring during the first months of 2008 and the
ARGO-YBJ experiment again reported evidence for a TeV emission in
correlation with the X-ray flares. The significance map of the
Mrk421 region is shown in Fig.\ref{fig:mrk421_08}: a clear signal
at about 7 standard deviations level is visible during the 2008
(days 1 - 229, 1217 hours on-source). The observation refers to a
multiplicity N$_{pad}\geq$60. A correlation of TeV photons
detected by the ARGO-YBJ experiment with the X-ray events detected
by the Rossi RXTE Satellite is evident in Fig.
\ref{fig_mrk421_xray} for two different multiplicity values.

%A correlation is evident, nevertheless deeper analysis is in
%progress to study the correlation between the X-ray flares and the
%TeV emission as a function of the photon energy.

We note that an all-sky VHE gamma-ray telescope as the ARGO-YBJ
experiment is able to monitor the Mrk421 in a continuous way, not
being affected by the problems of the Cherenkov telescopes which
can operate only on clear moonless nights.

%%%%%%%%%%%%%%%%%%%%%%%%%%%%%%%%%%%%%%%%%%%%%%%%%%%%%%%%%%%%%%%%%%%%
%%%%%%%%%%%%%%%%%%%%%%%%%%%%%%%%%%%%%%%%%%%%%%%%%%%%%%%%%%%%%%%%%%%%

\subsection{Search for GRBs with the Scaler Mode}

The Scaler Mode technique offers a unique tool to study GRBs in
the GeV energy range, where $\gamma$-rays are less affected by the
absorption due to pair production in the extragalactic space. In
order to extract the maximum information from the data, several
GRB searches have been implemented: (1) search in coincidence with
the satellite detection; (2) search for a delayed or anticipated
signal of fixed duration; (3) phase pile-up of all GRBs. The
search has been done in coincidence with 38 GRBs observed by
satellites (mainly SWIFT) in the period December 2004 - March 2008
in the ARGO-YBJ field of view ($\theta <45^{\circ}$). No deviation
from the statistical behaviour has been found either in
coincidence with the low energy detection, or in an interval of 2
hours around it (Di Girolamo T. et al., 2007). The corresponding
upper limits on the fluence are of the order of 10$^{-6}$ -
10$^{-4}$ erg/cm$^2$ in the 1 - 100 GeV range, well below the
energy range explored by the present generation of Cherenkov
telescopes.

\section{Cosmic Ray Physics}

\subsection{Measurement of the Proton-Air Cross Section}

A preliminary measurement of the proton-air cross section has been
performed by exploiting the attenuation of cosmic ray flux with
increasing zenith angles $\theta$ (i.e., atmospheric depths)
$\propto$ exp(-x$_0\cdot$ (sec$\theta$-1)/$\Lambda_{obs}$), where
x$_0$ is the vertical atmospheric depth of the detector and
$\Lambda_{obs}$ is the observed attenuation length of air showers,
related to the mean free path $\lambda_{int}$ of the primary
through the parameter K=$\Lambda_{obs}/\lambda_{int}$. Therefore,
with a measurement of the attenuation length we can estimate, in
principle, the p-air and p-p cross sections. The K parameter,
which takes into account the fluctuations both in the shower
development and in the shower sampling, is calculated via MC
simulation. The analysis has been performed in two multiplicity
bins, 300$\leq$N$_{pad}\leq$1000 and N$_{pad}>$1000, corresponding
to mean energies (3.9$\pm$0.1) TeV and (12.7$\pm$0.4) TeV,
respectively. The measured p-air cross section values are
(275$\pm$51) mb and (282$\pm$31) mb, respectively (De Mitri I. et
al., 2007) (see Fig.\ref{fig:pair}). Further studies are in
progress in order to extend the measurement to the PeV energy
region.

\subsection{Measurement of the antiproton/proton ratio}

\begin{figure}[t!]
\begin{minipage}[t]{.47\linewidth}
\begin{center}
\epsfysize=5.cm \hspace{0.5cm}
\epsfbox{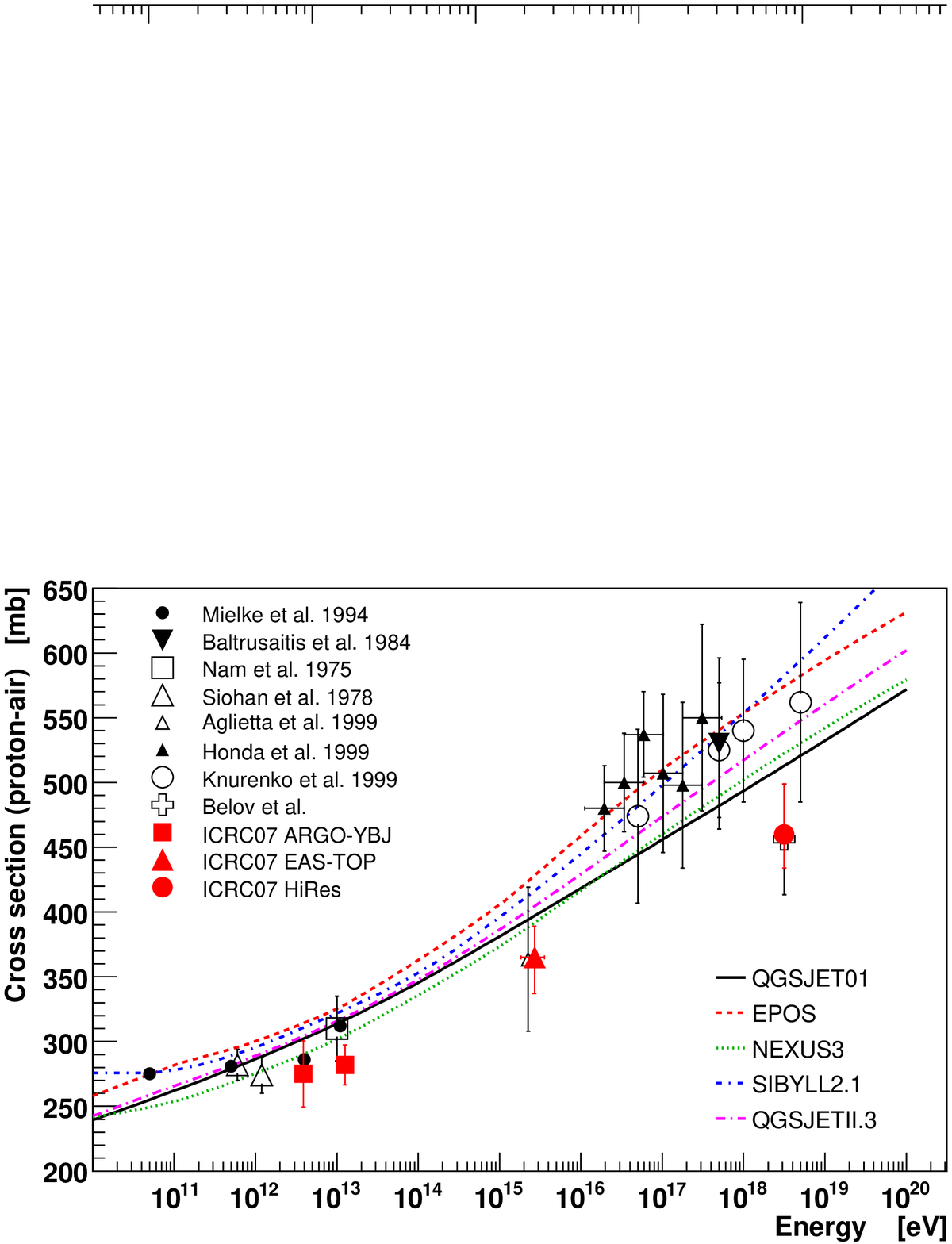} \vspace{-0.5cm}
\caption[h]{Current data of proton-air production cross section
measurements.} \label{fig:pair}
  \end{center}
\end{minipage}\hfill
\begin{minipage}[t]{.47\linewidth}
  \begin{center}
\epsfysize=5.cm \hspace{0.5cm}
\epsfbox{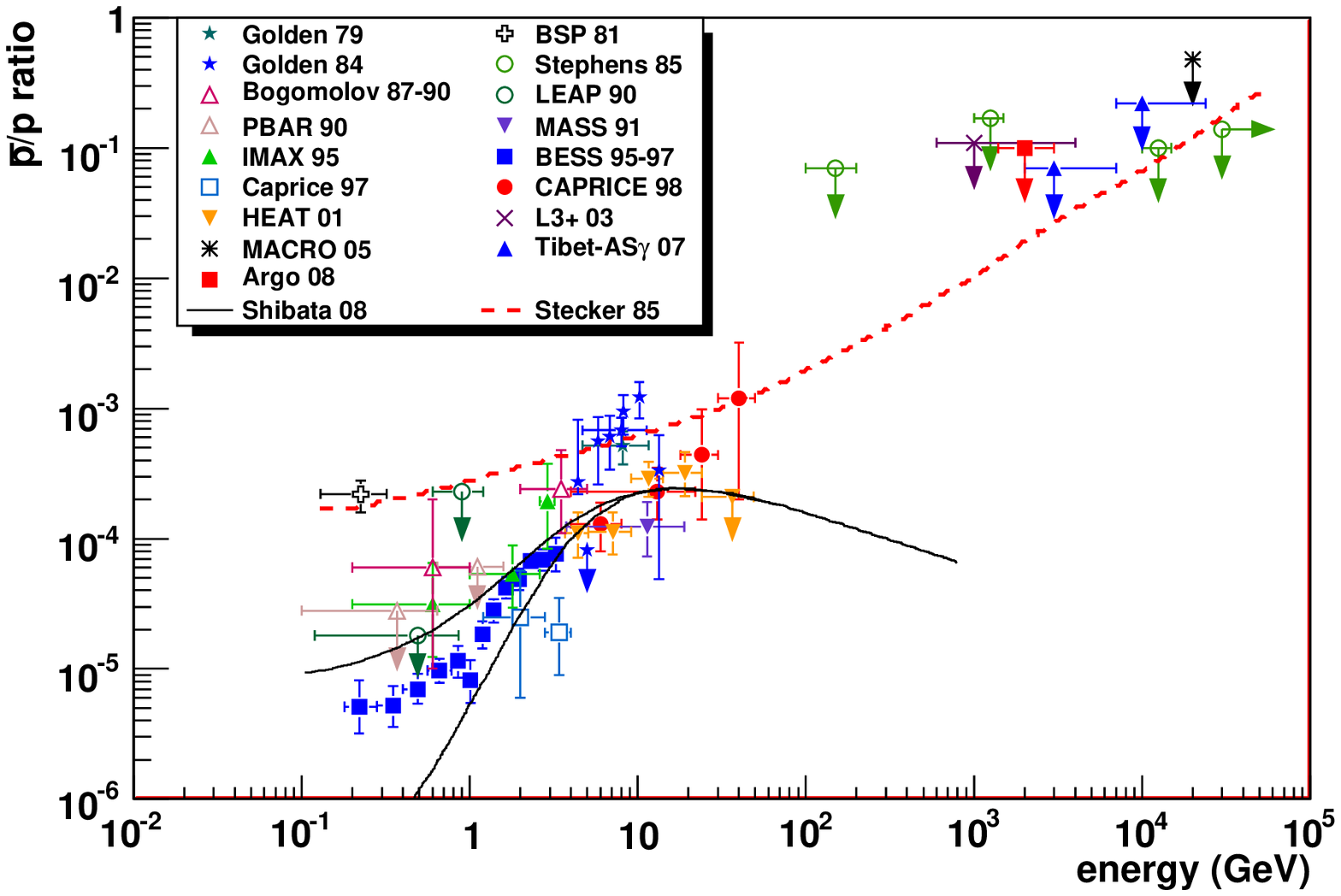} \vspace{-0.5cm}
\caption[h]{Status of the measurements of the $\overline{p}/p$
ratio.} \label{fig:antip}
  \end{center}
\end{minipage}\hfill
\end{figure}

In order to measure the $\overline{p}/p$ ratio at TeV energies we
exploit the Earth-Moon system as an ion spectrometer: if protons
are deflected towards East, antiprotons are deflected towards
West. If the energy is low enough and the angular resolution small
we can distinguish, in principle, between two shadows, one shifted
towards West due to the protons and the other shifted towards East
due to the antiprotons. If no event deficit is observed on the
antimatter side an upper limit on the antiproton content can be
calculated. A very preliminary measurement has been performed with
the ARGO-YBJ experiment for N$_{pad}\geq$40 in the period December
2007 - August 2008 (802 hours on-source). For this multiplicity
the Moon shadow shifts westward by about 0.45$^{\circ}$, at a
median energy $\approx$2 TeV (mode energy $\approx$0.5 TeV).
%The deficit events are equivalent to about 16 standard deviations
%on a two-dimensional event map.
The data is in fair agreement with MC simulation. The deficit
events around the Moon shadow
%(in the range $\pm$4$^{\circ}$ for the East-West direction and
%$\pm$2$^{\circ}$ for the North-South direction)
are fitted with a Gaussian formula: protons are estimated to be
$\sim$70\% of cosmic rays. A preliminary upper limit on the
antiproton/proton ratio at the 90\% confidence level is calculated
to be about 10$\%$ and reported in Fig. \ref{fig:antip} which
shows the status of the antiproton/proton measurements. We note
that in the multi-TeV range this result is among the lowest
available.

%%%%%%%%%%%%%%%%%%%%%%%%%%%%%%%%%%%%%%%%%%%%%%%%%%%%%%%%%%%%%%%%%%%%
%%%%%%%%%%%%%%%%%%%%%%%%%%%%%%%%%%%%%%%%%%%%%%%%%%%%%%%%%%%%%%%%%%%%

%%%%%%%%%%%%%%%%%%%%%%%%%%%%%%%%%%%%%%%%%%%%%%%%%%%%%%%%%%%%%%%%%%%%
%\begin{figure}[t!]
%\begin{center}
%%\special{psfile=figures/moon.eps  voffset=140 hoffset=0 hscale=30.0 vscale=18.7 angle=90}
%\epsfysize=6.5cm \hspace{1cm}
%\epsfbox{figures/disciascio_2008_01_fig09.eps} \vspace{-0.5cm}
%\caption[h]{Current data of proton-air production cross section
%measurements} \label{fig:pair}
%  \end{center}
%\end{figure}
%%%%%%%%%%%%%%%%%%%%%%%%%%%%%%%%%%%%%%%%%%%%%%%%%%%%%%%%%%%%%%%%%%%%

%\section{Conclusions}
%The ARGO-YBJ experiment completely installed has been in stable
%data taking since November 2007 in the YangBaJing Cosmic Ray
%Laboratory. Interesting results in Gamma-Ray Astronomy (Crab
%Nebula and Mrk421 observations, search for high energy tails of
%GRBs) and Cosmic Ray Physics (Moon and Sun shadow observations,
%proton-air cross section and antiproton/proton preliminary
%measurements) are already available and have been shortly resumed
%in this paper.

%\newpage
\bigskip
\bigskip
\noindent {\bf DISCUSSION}

\bigskip
\noindent {\bf A. SANTANGELO:} In the plots of the Crab appears
that a smaller significance is obtained with a longer exposure
time. Can you tell us why?

\bigskip
\noindent {\bf G. DI SCIASCIO:}  The statistical significance of
the 2008 Crab observation appears smaller than that of 2007. This
may have been caused by a statistical signal fluctuation: the 2007
data sample is about a factor of 5 smaller than that integrated up
to August 2008. Nevertheless this analysis is yet preliminary and
deeper reanalysis is in progress in order to evaluate the final
sensitivity of the detector properly.

\end{document}